\documentstyle[prl,aps,psfig,epsfig,multicol]{revtex}
\begin{document}
\draft 
\title{Is Li$_2$Pd$_3$B a self-doped hole superconductor ?}
\author{Manas Sardar$^1$ and Debanand Sa$^2$}
\address{$^1$ Material Science Division, Indira Gandhi Centre for 
Atomic Research, Kalpakkam-603 102, India \\ 
$^2$ The Institute of Mathematical Sciences, C. I. T. Campus,
Chennai-600 113}
\maketitle

\begin{abstract}  

We propose that the electrons responsible for superconductivity 
in Li$_2$Pd$_3$B come from the palladium 4d-electrons.
So, its electronic properties are likely to be dominated by strong
electronic correlations. The basic unit in this material are Pd$_6$B 
octahedra which share vertices to form a 3-dimensional network. Due 
to the highly distorted nature of the Pd$_6$B octahedron, one far 
stretched Pd atom per octahedra becomes almost inactive for electronic 
conduction. Thus, the material escapes the fate of becoming a 
half-filled insulating Mott antiferromagnet by hiding extra charges at 
these inactive Pd sites and becomes a self-doped correlated metal.   
We propose a 3-dimensional single band t-J model which could be the 
correct minimal model for this material. 

\end{abstract} 

\begin{multicols}{2} 

\vskip 1cm
\narrowtext 

Recent discovery of 8 K superconductivity in a non-oxide material 
Li$_2$Pd$_3$B \cite{togano04},  
has once again opened up challenges for
search of newer superconductors. This is the first observation of
superconductivity in metal rich ternary borides
containing alkaline metal and Pd ( Pt group), a 4d transition metal.

First metal boride to show superconductivity (T$_c$=4 K) was TaB$_2$,  
discovered by Kiessling \cite{kies49}. Since then many binary and
ternary superconducting borides involving 3d transition and rare earth
metals were discovered. For a complete bibliography see \cite{buzea01}.

T$_c$ remained below 12 K before the discovery of MgB$_2$ (T$_c$=39 K) 
by Nagamatsu ${\it et\>\> al.}$, \cite{nagamatsu01}.
It still remains a surprise that how such high T$_c$ in this material 
can be obtained, even though the present view is that, MgB$_2$ is a 
phonon mediated strong coupling superconductor.

For Li$_2$Pd$_3$B compound, transport and spectroscopic properties are
still unexplored, even though the detail structural data is
available \cite{eib97}.

In this short communication, we propose that the electrons responsible
for superconductivity in Li$_2$Pd$_3$B comes from the palladium 
4d-electrons.
So, its electronic properties are likely to be dominated by strong
electronic correlations. The basic unit in this material are Pd$_6$B 
octahedra which share vertices to form a 3-dimensional network. Due 
to the highly distorted nature of the Pd$_6$B octahedron, one far 
stretched Pd atom/octahedra becomes inactive for electronic conduction. 
Thus, {\em the material escapes the fate of becoming a half-filled  
insulating Mott antiferromagnet by hiding extra charges at 
these inactive Pd sites and becomes a self-doped correlated metal.   
This is a classic example of Mott insulator doping itself at the 
cost of elastic energy by deforming the octahedral cages}. 
We propose a 3-dimensional single band t-J model which could be the 
correct minimal model for this material. 

The structure of Li$_2$Pd$_3$B is cubic having the space group 
P4$_3$32. This is similar to Li$_2$Pt$_3$B. 
The characteristic building blocks are distorted
Pd$_6$B octahedra, centered by boron atom. This is similar to metal
oxide superconductors, such as, high T$_c$ oxides, barium bismuthates, 
strontium ruthenetes and sodium cobaltates etc.. The difference being,
in oxide superconductors, 
metal atoms are enclosed by oxygen octahedral cage, whereas
in the present material, boron is octahedrally coordinated by Pd metal
ions. The Pd  octahedra share vertices and form a 3-dimensional network.
Lithium partial structure is also interesting. 
Every Li atom has 3 neighbouring
Li atoms at a distance of 2.55~\AA. Li-Li distance in Li metal is about
3.05~\AA. Since Li-Li distance in this material is about 16$\%$ smaller
compared to lithium metal, one can assume that there is complete
transfer of electrons from lithium to Pd-B complex. There are no 
close contacts between lithium and boron atoms.  
 
As every palladium atom forms the common vertex of two Pd$_6$ octahedra, 
there are eight neighbouring Pd atoms. Out of four Pd atoms in the 
basal plane above, one is pulled too far out, so that the four Pd-Pd 
distances are $2 \times$ 2.78~\AA, 2.95~\AA, 3.52~\AA.
Similar stretching occurs in the basal plane below also. The stretched
Pd atoms in  the planes below and above are in different directions 
(See Fig. 1).
Ultimately,  one has four short Pd-Pd distance (2.78~\AA), two medium
bonds (2.95~\AA) and two very long bonds (3.52~\AA).
The  short and the medium Pd-Pd bonds (total six in number)
are slightly larger than metallic palladium, but the long one is
too large for binding interaction (Fig. 1). 

Boron is the group 13 element in the periodic table 
with atomic number 5. Its electronic configuration is 
1s$^2$, 2s$^2$, 2p$^1$. This means, it can easily accept electrons 
from the d-shell when it tries to form complexes with other elements 
such as Pd, Pt etc.. It is also known from the elementary chemistry 
that boron has the property of distorting a structure whenever it forms 
metal complexes, hydrides etc. That is the reason why one observes a 
varieties of structures in boron based materials \cite{knoth68}.     

\begin{figure}
\begin{center} 
\psfig{figure=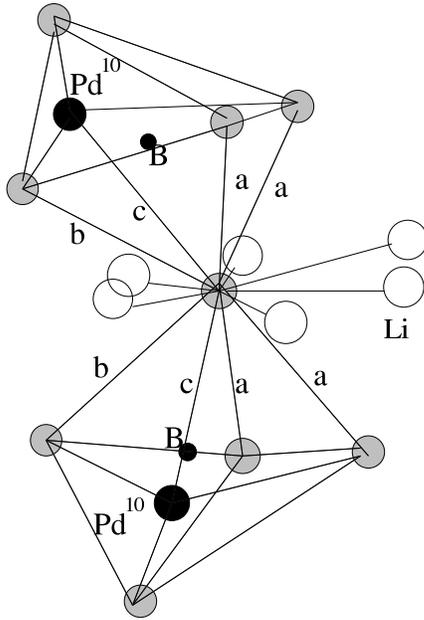,width=0.65\linewidth}
\caption{Two vertex shared Pd$_6$B octahedra where Pd-Pd bond 
length `c' is the largest as compared to `a' and `b'. The inactive 
Pd atoms which don't take part in electronic activities 
are shown as dark large circles. Also the Boron 
(dark small circles) and Lithium (large open circles) atoms are 
shown.  \label{li1}}
\end{center} 
\end{figure}

\begin{figure}
\begin{center} 
\psfig{figure=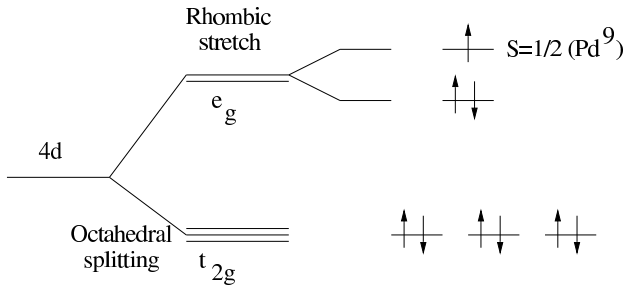,width=0.95\linewidth}
\caption{Crystal-field split of Pd 4d levels.  \label{li2}}
\end{center} 
\end{figure}

From nominal valence counting in Li$_2$Pd$_3$B, boron will be in (-5)
valence state to make a closed 2p$^6$ configuration. Since lithium is in
(+1) valence state, the average Pd valency becomes +1 (4d$^9$), which 
implies, one electron/hole per Pd atom. Since the material has a   
3-dimensional cubic perovskite structure, the 4d level splits into 
t$_{2g}$ and e$_{g}$ levels due to octahedral coordination, 
out of which t$_{2g}$ will be completely 
filled (see Fig. 2). Due to unusual rhombic stretch of the octahedra,  
Pd acquires a very low site symmetry. This makes e$_{g}$ levels to 
split further and out of remaining 3 electrons, 2  occupy the lower 
level leaving the last one in the topmost e$_{g}$ level. 
This  gives rise to one electron/Pd site, i.e., 
a half-filled non-degenerate single band situation. 
It should be noted that the the Pd$_6$B octahedra with 
rhombic stretch has escaped the important effects such as 
Jahn-Teller distortion, Hund coupling and possible high spin 
ground state of Pd ions which would have worked against 
superconductivity. At this stage, a comparison with another 
non-oxide superconductor MgNi$_3$C is called for
since it has close structural similarity.

MgNi$_3$C has a very symmetrical cubic perovskite structure at high 
temperature \cite{huang01} 
with vertex sharing nickel octahedra enclosing carbon. This is symmetric in
all 3 directions, giving rise to isotropic hopping integral in a tight
binding sense. The band width estimated from density functional
theory \cite{singh01} is about 2.5 eV, i.e., hopping `t' is 
about 0.2 eV.
We believe that the hopping integral for the present material will be even
smaller than that of MgNi$_3$C, since the structure here is badly distorted.
The most important difference in Li$_2$Pd$_3$B as compared to MgNi$_3$C 
is one far stretched ({\em lone}) Pd site/octahedra due to distortion. 
The hopping matrix element connecting this {\em lone} Pd site 
to the other ones will be vanishingly small. This will  make the 
rest of the conducting Pd backbone look effectively very different 
than the 
original network of octahedra, but would still form a 3-dimensional 
network due to large coordination number of Pd atoms. 
Moreover, due to the 4d character of the carriers, the local Coulomb 
correlation (U) in this material is expected to be large, similar 
to that of MgNi$_3$C, i.e., $\sim$ 4 eV. 
Thus, in the limit of strong local Coulomb correlation,  
one has a Mott insulator and the correct modeling will
be a spin-half Heisenberg antiferromagnet on a 3-dimensional lattice.
As in other transition metal oxides with a strong Hubbard repulsion, 
it will lead to the usual superexchange integral between two Pd ions,  
but since the paths here are not 180$^0$, the antiferromagnetic coupling 
will be reduced in strength. Considering the above parameters for 
`t' as well as `U', we estimate the value of J to be   
$\sim$ 50 meV, i.e., nearly 500 K. Magnetic susceptibility 
measurement along with a careful tight binding analysis will be able 
to confirm these estimates.   

The {\em lone} Pd site (one per octahedra), 
will retain its closed shell electronic configuration 4d$^{10}$ 
and won't interact with the rest of Pd  d-electron spins. As already been 
mentioned earlier, due to the inactive nature of the {\em lone}   
Pd atom in the octahedra, the average occupancy will not be 9.0 as one 
would have naively expected. Since the {\em lone} Pd site/octahedra 
( ${1\over 6}$-th of the total palladium sites) holds back the 
electronic configuration 4d$^{10}$ (atomic), the rest ${5\over 6}$-th 
of the palladium atoms will have average occupancy 
of 8.8 electrons/Pd site, i.e., 20$\%$ hole doping. 
Thus, it becomes a self-doped system  over a Mott antiferromagnetic 
insulating background on a leftover 3-dimensional lattice (where the 
average Pd-Pd coordination number becomes 7 instead of 8 if it would 
have been undistorted Pd$_6$B octahedra). For such a situation, the 
minimal model would be a 3-dimensional single band t-J model. This 
can be contrasted with the earlier proposal of self-doping due to 
externally applied pressure/chemical pressure in transition 
metal oxide superconductors, fullerides as well as in organics 
\cite{baskaran03}. In the present case,  {\em self-doping is done 
self-consistently by the system itself at the cost of a  
peculiar distortion of the Pd octahedra}.  
    
Since, our modeling indicates that the carriers in Li-Pd-B ternary 
borides are holes, one would expect a positive sign in Hall 
as well as thermopower measurements at low temperature (above T$_c$) 
which can be tested experimentally. 

The next question is then, why would it need such a large 
(20$\%$ doping) carrier concentration to metallize and ultimately 
give rise to superconductivity? The answer would be the following: 
If the {\em lone} Pd atoms would have taken part in the electronic 
activity, then one would have a perfect 3-dimensional Heisenberg 
antiferromagnet with possible long range order. In order to destroy 
a 3-dimensional antiferromagnetic long range ordering and to metallize 
it (as compared to usual 2-dimensional systems),  one would need to 
go for much higher doping. This dimensional argument might be the reason 
why one needs more doping for this material than that of standard 
2-dimensional (optimal hole doping in cuprate superconductors is about 
15$\%$) materials. 

However, it should be noted that in the process of making one 
Pd atom/octahedra electronically inactive, the system looses 
some of its elastic energy due to lattice distortion but gains 
electronic energy due to carrier delocalization. 
Since one observes superconductivity in the system, it implies that 
the electronic energy wins over the elastic energy.   

Much of the experimental data on Li-Pd-B ternary superconductor is 
unknown. The resistivity at 100 K, on the other hand, 
is too high compared to other metallic borides superconductors 
like MgB$_2$ ($\rho \approx 5 \> \mu\Omega$ cm) \cite{eltsev02},   
but it compares well  with other strongly 
correlated metallic superconductors like, layered cuprates, 
barium bismuthates, sodium cobaltates etc. (where $\rho \approx  
100 \>\mu\Omega$ cm at T= 100 K is typical \cite{iye92}). For comparison, 
an almost isostructural compound like MgNi$_3$C has  $\rho$(T=10 K) 
=125 $\>\mu\Omega$ cm and  $\rho$(T=300 K)=325 $\>\mu\Omega$ cm \cite{he01}. 
For Li$_2$Pd$_3$B \cite{bharathi04}, $\rho$(T=10 K)
=70 $\>\mu\Omega$ cm and  $\rho$(T=300 K)=120 $\>\mu\Omega$ cm. 
In case of  MgNi$_3$C, resistivity is linear in temperature upto 60 K 
as well as for T $>$ 60 K (up to 300 K) with a break of slope at T=60 K 
\cite{he01}. 
Similar change of slope in resistivity at 60 K has been observed for  
Li$_2$Pd$_3$B \cite{bharathi04}. The overall T-dependence of 
resistivity in this material is almost identical to that of 
MgNi$_3$C. The fall in resistivity at T=50 to 60 K suggests a 
pseudogap physics (reduction in density of states).  
Similarity with the physics of superexchange dominated  
strongly correlated metals such as high T$_c$ oxide is obvious. 

From the resistivity of $\rho=120\> \mu\Omega$ cm at T=300 K, 
we estimate the mean free path to be about 200~\AA, i.e., 30 lattice 
spacing, implying a good metal. One difference with cuprates is 
the large residual resistivity, $\rho\approx 50\> \mu\Omega$ cm. 
This is true for MgNi$_3$C as well. 
In Li$_2$Pd$_3$B, the {\em lone} defect Pd sites (non-conducting) 
might form a background charge density wave. The additional low energy 
scattering from the fluctuations of this charge density, would  
be responsible for such a large residual resistivity. We believe, this 
could be a generic feature of strongly correlated metals, where various 
kinds of density waves competes with superconductivity. 
This clearly indicates that electronic correlations in this 
material is at work which might be responsible for largely linear T 
dependence of low temperature resistivity. 
We believe that such non-Fermi liquid 
behaviour might continue to occur in many other 
measurements also. 

The basic degrees of freedom and their interactions can be captured 
by a model hamiltonian which can be written as, 

\begin{eqnarray} 
H = -t \sum_{<i,j>\epsilon \{A\}, \sigma}
(C_{i\sigma}^\dagger C_{j\sigma} + h. c.) 
+J\sum_{<i,j>\epsilon \{A\}} {\bf S_i}\cdot {\bf S_j} \nonumber \\
-E_I\sum_{j\epsilon \{A\}, \sigma} n_{I\sigma}{\bf u_{Ij}}  
+{\lambda\over 2}\sum_{i\epsilon \{B\}, j\epsilon \{A\}}{\bf u_{ij}}^2 
\end{eqnarray} 

\noindent where, `t' and `J' are the standard hopping and the 
superexchange integral as has already been mentioned. 
Here, $C_{i\sigma}^\dagger $ is the electron creation operator 
at site `i', with spin projection `$\sigma$' and ${\bf S_i}$, the 
spin operator of palladium 4d$^9$ active sites.
$\{A\}$ denotes sublattice A for active Pd sites (${5\over 6}$-th of total 
Pd sites) and $\{B\}$ refers to the rest 
(inactive {\em lone} Pd sites). $E_I$ is 
the reduction (note the negative sign) in the inactive Pd site energy 
(I is the inactive Pd site) due to local distortion of the lattice, 
i.e., whenever $\sum_j<{\bf u_{Ij}}> \ne 0$, ${\bf u_{ij}}$ 
being the lattice displacement vector between i and j Pd atoms. 
The last term in the hamiltonian is the increase in the elastic energy 
cost due to lattice distortion and  $\lambda$, the elastic constant.   
The last two terms in the hamiltonian represent the balance between 
the electronic and the elastic energies. 

A standard RVB mean field theory for the above 3-dimensional t-J model 
will provide a rough estimate for the superconducting transition 
temperature (T$_c$), $k_BT_c \approx  {W\over 2} e^{-{{W}\over {J}}}$. 
Assuming the bare band width W to be nearly 1.5 eV (It is reduced as  
compared to MgNi$_3$C because of the highly distorted nature of the 
present material), 20$\%$ doping makes it even smaller, i.e., 0.3 eV. 
Considering J to be nearly 50 meV, one obtains T$_c$ $\approx $ (1-10) K, 
which compares well with the experiment.  

Superconductivity in Li-Pd-B ternary boride has been confirmed from 
magnetization as well as resistivity measurements. Magnetoresistivity 
measurement \cite{togano04,badica04} reveals that the upper 
critical field H$_{c2}$(0) in this 
material is about 4 T which is well within the Pauli paramagnetic 
limit. Considering this value of H$_{c2}$(0), one can estimate the 
coherence length $\xi$(0) which turns out be much larger 
(few hundred ~\AA) than the standard high T$_c$ superconductors.   
It is well known that in Sr$_2$RuO$_4$ \cite{maeno03}, 
which is a p-wave (triplet) 
superconductor, even a small disorder giving rise to a residual 
resistivity of 1.5 $\>\mu\Omega$ cm, is capable of suppressing 
superconducting T$_c$ to zero. The large residual resistivity 
along with a low  H$_{c2}$(0) (no Pauli paramagnetic limit for 
H$_{c2}$(0) exists for triplet superconductors), possibly disfavours 
a triplet pairing and thus, a likely candidate could be a singlet one 
in this material. 

To conclude, we have analyzed the structural data of a recently discovered 
Li-Pd-B ternary superconductor. From the elementary electron counting 
combined with the structural data, we arrive at the conclusion that the 
low temperature electronic properties in this material are dominated by 
strong electronic correlations. The basic units in this material 
are highly distorted Pd$_6$B octahedra, which share vertices to form a 
3-dimensional structure. Due to the distorted nature, one far stretched 
Pd atom/octahedra becomes electronically inactive and hence the material 
becomes self-doped with 20$\%$ holes as the carriers. The 4d nature 
of the Pd valence electrons are known to have large on-site Coulomb 
repulsion and the appropriate model would be a 3-dimensional single 
band t-J model. {\em This could be the first example of 
realising superconductivity 
in a 3-dimensional t-J model}. Even though most of 
the transport and spectroscopic properties in this material are yet 
to be explored, we expect them to exhibit non-Fermi liquid behaviour 
at low temperature. More experiments are needed to understand the 
similarities as well as the differences of all the normal state properties 
of a predominantly 3-dimensional compound like Li$_2$Pd$_3$B and 
other quasi two 
dimensional superconductors such as high T$_c$ oxides, organics, sodium 
cobaltates etc.. Existence of superconductivity in 2-dimensional t-J 
model on a square lattice, is no longer doubted \cite{bas87,param02}. 
The search for superconducting ground states in 3-dimensional t-J model 
on various kinds of lattices need to be explored.      
 
It is a pleasure to thank G. Baskaran for useful discussions 
and bringing out Ref.[6] to our attention. We also thank R. Shankar 
and V. Subrahmanyam for discussions.

\end{multicols}
\end{document}